\documentclass [12pt,twoside]{article}           
\usepackage[left,modulo]{lineno}
\usepackage{setspace}

\countdef\arxiv=201

\ifnum\arxiv=0 \input cpreb.tex \fi

\ifnum\arxiv=1

\usepackage{epsfig,times,lscape}
\usepackage[usenames]{color}

    \ifnum\count102=1

    \fi

    \ifnum\count102=2
\fi

\newcommand{\nc}{\newcommand}

     \ifnum\count101=1
\nc{\qI}[1]{\section{{#1}}}
\nc{\qA}[1]{\subsection{{#1}}}
\nc{\qun}[1]{\subsubsection{{#1}}}
\nc{\qa}[1]{\paragraph{{#1}}}

\def\qpar{\vskip 2mm plus 0.2mm minus 0.2mm}
\def\qL{\hfill \break}
     \fi 

      \ifnum\count101=2
 \nc{\qI}[1]{\parindent=0mm \vskip 8mm 
{\centerline{\LARGE \color{red}#1}}\vskip 3mm}
\nc{\qA}[1]{\vskip 2.5mm \noindent 
{{\bf\large\color{blue}  #1}} \vskip 1mm \parindent=0mm}
 \nc{\qun}[1]{\vskip 1mm \noindent {\sl #1 }\quad }

\def\qL{\hfill \break}
\def\qpar{\vskip 2mm plus 0.2mm minus 0.2mm}

      \fi

\nc{\qfoot}[1]{\footnote{{#1}}}

      \ifnum\count101=1
\def\qbu{\hfill \par \hskip 6mm $ \bullet $ \hskip 2mm}
\def\qee#1{\hfill \par \hskip 6mm (#1) \hskip 2 mm}
      \fi
      \ifnum\count101=2
\def\qbu{\hfill \par \hskip 4mm $ \bullet $ \hskip 2mm}
\def\qee#1{\hfill \par \hskip 4mm (#1) \hskip 2 mm}
      \fi

\def\qparr{ \vskip 1.0mm plus 0.2mm minus 0.2mm \hangindent=10mm
\hangafter=1}

     \ifnum\count101=1 
 \def\qdec#1{\parindent=0mm\par {\leftskip=2cm {#1} \par}}
     \fi
     \ifnum\count101=2
  \def\qdec#1{\parindent=0mm \par {\leftskip=1cm {#1} \par}}
  
  \def\qcitb#1{\noindent \hbox to 102mm{\hfill \small #1} \vskip 1mm}
      \fi

 \def\qpages#1{\count102=0{\loop\advance\count102 by 1
 \null \vfill\eject \ifnum\count102<#1 \repeat}}

\def\qv{\vskip 0.1mm plus 0.05mm minus 0.05mm}
\def\qhu{\hskip 0.6mm}
\def\qhv{\hskip 3mm}

\def\qhw{\hskip 1.5mm}
\def\qleg#1#2#3{\noindent {\bf \small #1\qhw}{\small #2\qhw}{\it \small #3}\qv }
\fi

\begin{document}
\thispagestyle{empty}



\markboth{{\sl \hfill  \hfill \protect\phantom{3}}}
        {{\protect\phantom{3}\sl \hfill  \hfill}}

\color{yellow} 
\hrule height 10mm depth 10mm width 170mm 
\color{black}

\vskip -12mm 

\centerline{\bf \Large Similarities vs. key discrepancy between
  tuberculosis and cancer}
\vskip 8mm

\centerline{\large 
Peter Richmond$ ^1 $ and Bertrand M. Roehner$ ^2 $
}

\vskip 5mm
\large

%
{\bf Abstract}\qL
In 2015 in the United States 612,000 persons died
from cancer whereas 
only 470 died from tuberculosis (TB), a disease which was the
main cause of death around 1900. How can one
explain such a key discrepancy in treatment progress?
A statistical comparison between TB and cancer will 
give some clues.\qL
However, TB and cancer also share several important features.
Both TB and cancer can affect several organs,
e.g. lungs, brain, bones, intestines, skin. 
What in cancer is called malignant neoplasm (tumor) 
is called granuloma in TB.
By isolating
malignant cells (versus bacteria) from the rest of the body,
such clusters protect the host's organism but at the same
time they are ``secure beachheads'' from where malignant cells
(versus infected macrophages) can wander off to new locations.
Thus, metastatic tumors have a TB parallel in the form of
secondary granulomas.\qL
In order to investigate more closely this parallel
we use 
the age-specific response of organs.
Called spectrometric analysis in a previous
paper (Berrut et al. 2017), this method
provides information about how fast tumors develop and
how serious they become. 
A convenient characterization
of the response to TB of organ $ j $ is 
given by the following (age-dependent) death ratio:\qL
\centerline{
$ T_j(t)=$ ({\small death by TB of type $ j $})/({\small all TB deaths}).}\qL
The development of cancer tumors can be described by 
a similar function
$ C_j(t) $.\qL
We compare the organs' responses
in all cases for which specific death data are available.
It appears that for the same organ
$ T_j(t) $ is similar in shape to $ C_j(t)  $. 
In other words, the idiosyncrasies of each organ are
more determinant than the functional differences
of TB versus cancer.
For instance, with regard to brain tumors,
both TB and cancer mortality peak around the age of 10. \qL
Such observations  
may bring to light vulnerabilities in the way
the immune system provides protection to various organs.

\vskip 3mm
\centerline{\it \small Provisional. Version of 13 May 2018. 
Comments are welcome.}
\vskip 3mm

{\small Key-words: age-specific death rate, death ratio, 
tuberculosis, cancer}

\vskip 3mm

{\normalsize
1: School of Physics, Trinity College Dublin, Ireland.
Email: peter\_richmond@ymail.com \qL
2: Institute for Theoretical and High Energy Physics (LPTHE),
University Pierre and Marie Curie, Paris, France. 
Email: roehner@lpthe.jussieu.fr
}

\vfill\eject

\qI{Introduction}

\qA{Contrasting death rates}

In the early 20th century in Europe and the United States
tuberculosis was the first cause of death.
Cancer came only in fourth position after cardiovascular
and cerebrovascular diseases.
Nowadays, although remaining a substantial cause
of death in old age, tuberculosis 
has been virtually eliminated in young and middle-aged
patients. In contrast, despite intensive research,
progress in cancer treatment was much slower.

%
\begin{figure}[htb]
\centerline{\psfig{width=17cm,figure=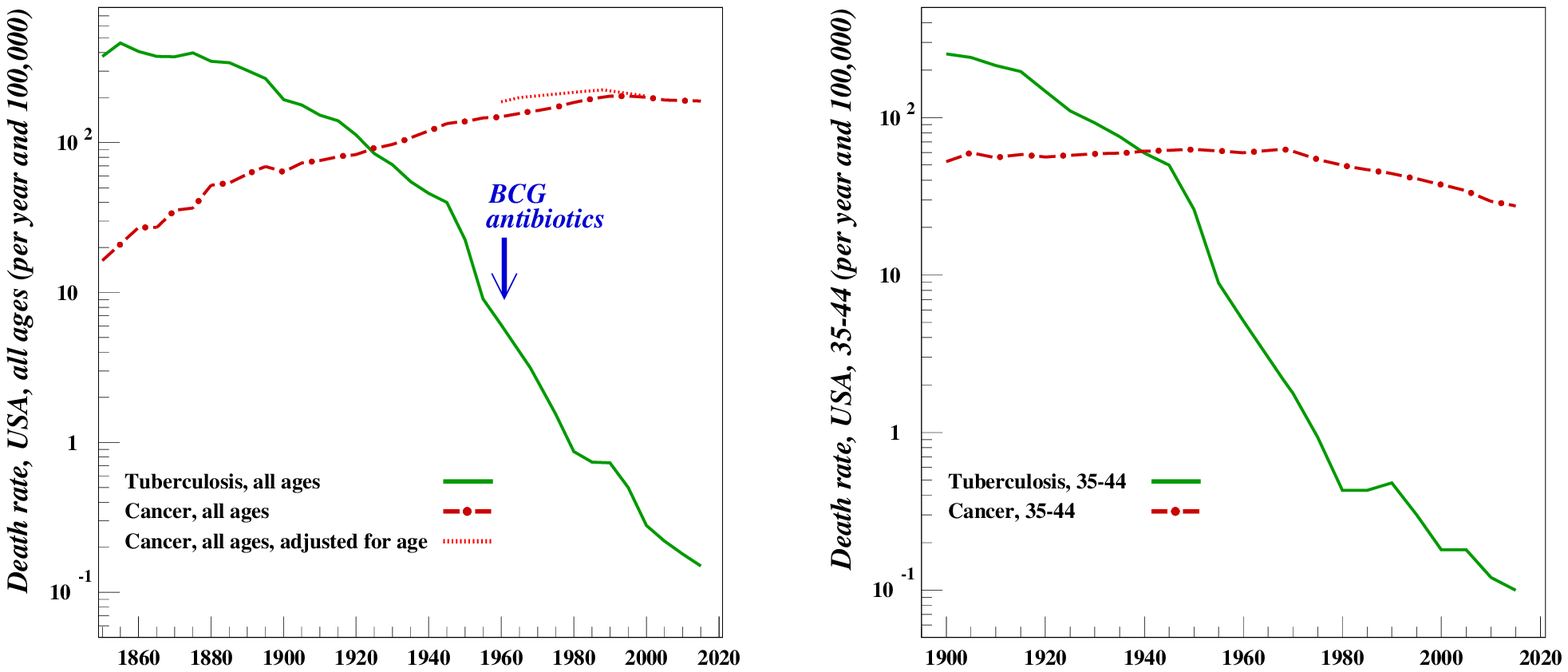}}
\qleg{Fig.\qhu 1a,b\qhv Comparison between the death rates
of TB and cancer in the United States, 1900-2015}
{The TB death rate was divided by 17 even before
the introduction of the BCG vaccine and of antibiotics
in the 1950s.
Note that age-adjustment that takes
into account the increase in the number of elderly people
makes in fact little difference.}
{Sources: 1850--1899: Statistique internationale du mouvement de la
population (1907);
1900--1940: Linder et al. (1947); 
1940--1960: Grove et al. (1968); 1968--2015 : 
CDC Wonder databases, compressed mortality
1968--1978,1979--1998,1999--2016. 
1950-2002 (adjusted for age): National Vital Statistics
Report 50,15,September 16, 2002.
The respective
ICD (International Classification of Diseases) codes for
TB and cancer are as follows. 
1968-1998: ICD8 and ICD9: 010-019, 140-239, 
1999-2016: ICD10: A16-A19, C00-D48.}
\end{figure}

Fig. 1a,b shows a dramatic difference between the death rate
changes of the two diseases. How can one possibly explain it?

%
\begin{figure}[htb]
\centerline{\psfig{width=15cm,figure=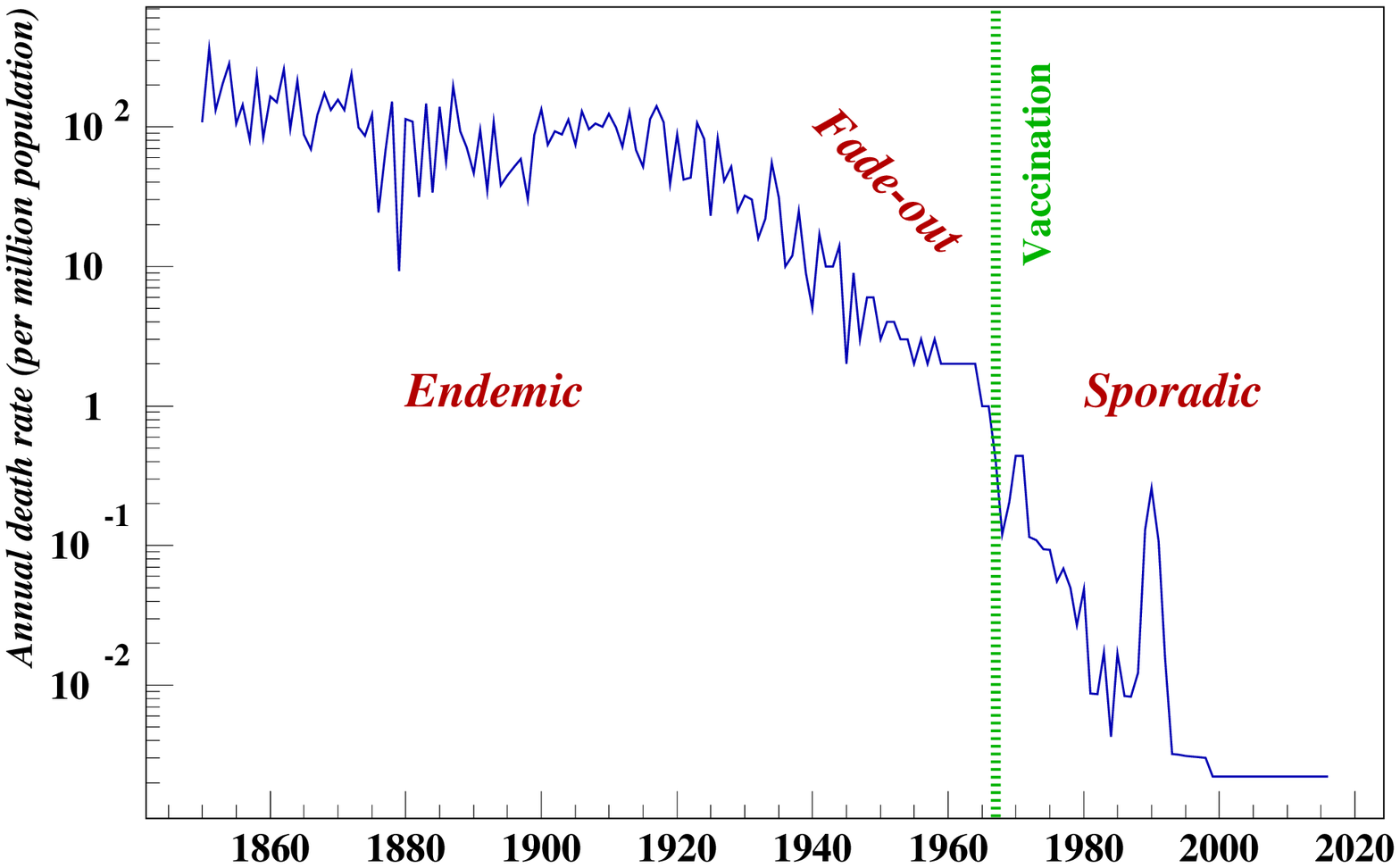}}
\qleg{Fig.\qhu 1c\qhv Natural history of the death rate 
of measles in the United States.}
{For 1850--1899 the data are for the state of Massachusetts
which is the only state to have data going back to 1850.
For 1900--1930 the data are for the registration states.
After 1930 the data cover the whole country (except
Alaska and Hawaii until statehood). In 1989--1991
there was a measles outbreak which particularly affected
California, Chicago and New York.}
{Sources: 
1850--1899: Statistique internationale du mouvement de la
population (1907);
1900--1940: Linder et al. (1947); 
1940--1960: Grove et al. (1968); 
1961--1967: Mortality statistics in the United States 1967 p.1-7;
1968--2015 : CDC Wonder databases, compressed mortality
                      1968--1978,1979--1998,1999--2016. 
The respective
ICD (International Classification of Diseases) codes for
measles are as follows. 
1968-1998: ICD8 and ICD9: 055, 
1999-2016: ICD10: B05.}
\end{figure}

Before we develop our explanation let us summarize the key points 
in a few sentences.
\qbu For cancer the incidence rate (i.e. the annual number of
new cases in a population of 100,000) is 439 whereas the death rate
is 163%
\qfoot{These data are given by the US National Cancer Institute for
the years 2011--2015.}%
.
This gives a case-fatality ratio (i.e. number of fatalities for
100 persons who became ill) of $ 163/439=37\% $.
This is a proportion which is higher than for most infectious
diseases. Thus, for measles in the outbreak of 1989--1991 in
California and New York it was less than 1\%. As a more
significant comparison, the case fatality ratio of TB for
South African HIV-infested persons is about 20\% (Walt et al. 2016).
Moreover, for most infectious diseases the case-fatality ratio
decreases with age whereas for cancer it increases with age
\qfoot{For instance, it is reported by
Dales et al. (1993) that for measles during the outbreak of
1989--1991 in California and New York City, 
the case fatality
rate was divided by a factor 100 between the age intervals
(0,1\ year) and (5\ y,15\ y).}.
On the contrary, for cancer the case fatality ratio increases
with age (Fig. 1d). 
%
\begin{figure}[htb]
\centerline{\psfig{width=10cm,figure=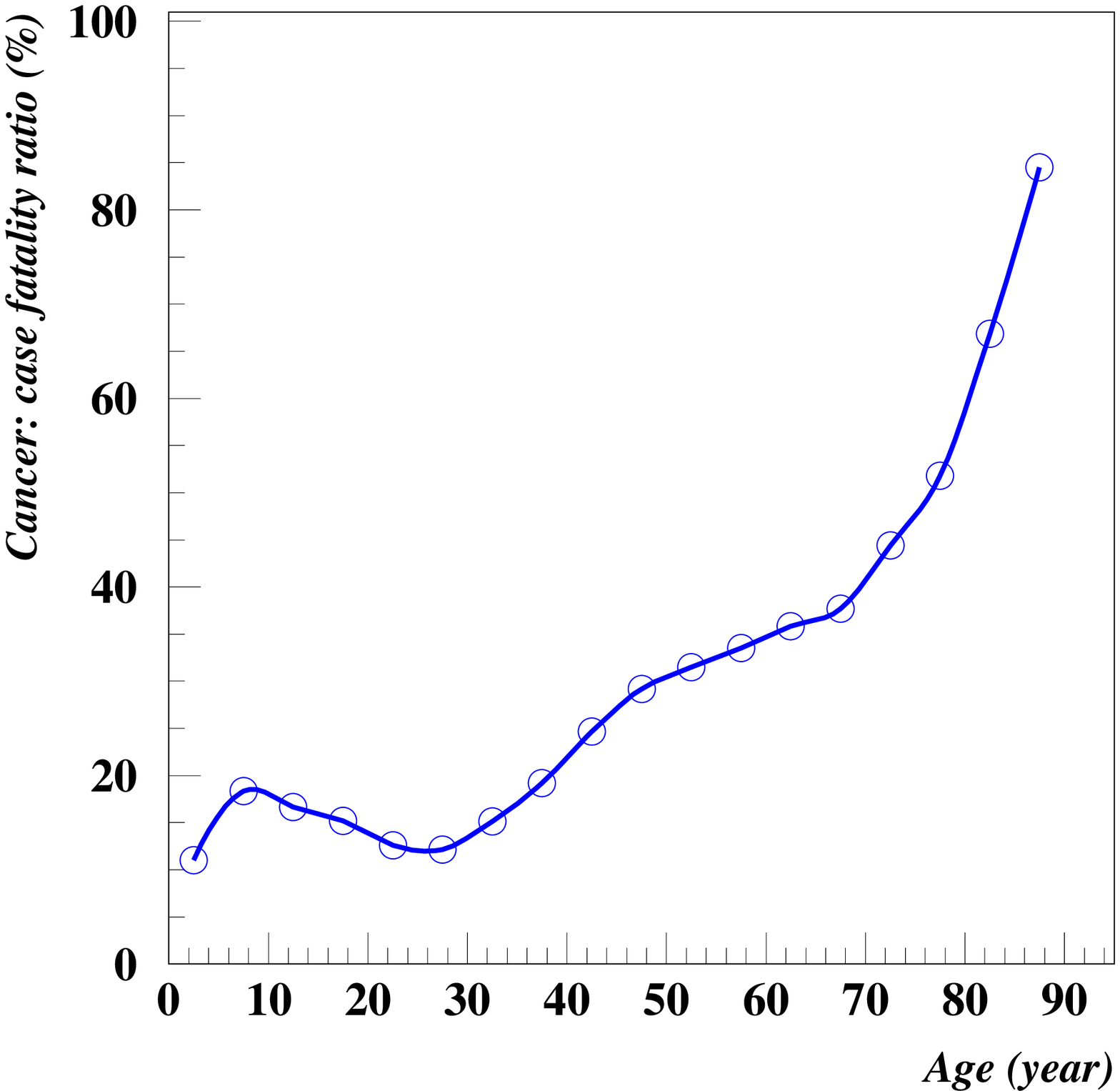}}
\qleg{Fig.\qhu 1d\qhv Cancer: case fatality ratio.}
{The case fatality ratio is the ratio: death rate divided
by incidence rate. It gives the probability of death 
for a person who becomes ill in the respective age interval.
The fact that the ratio increases with age shows that the
organism becomes less and less able to fight the disease
successfully. In contrast, for most infectious 
disease the CF ratio decreases with age.}
{Source: Cancer Research UK. The data are for all forms
of cancer (C00-C97), 2013-2015.}
\end{figure}
%
\qbu In an epidemic usually only a small proportion of the
population is infected by the pathogens in the sense that the
others do not produce any antibody. On the contrary, the
production of abnormal cells occurs in everyone which means that
the infection rate is 100\%.
\qbu With respect to infectious diseases any population
comprises two components: (i) the susceptibles that is to say the
persons who are not immunized and (ii) those who are immunized
and the proportion of the second group increases (i) with age
and (ii) for each age-group in the course of time, at least
in the fade-out phase.
On the
contrary, for cancer the ability to fight the disease
decreases with age as shown by Fig. 1d and does not improve
in the course of time.
\qpar

In summary, the previous points together with the upcoming discussion
suggest that cancer is fundamentally different from
infectious diseases and that human organisms are not
well equipped to fight it. Hence the huge difference displayed
in Fig. 1a,b.

\qA{Natural history of infectious diseases}

The history of infectious diseases like smallpox, measles or TB
usually comprises 4 phases, each of which may last from
a few decades to a few centuries. Despite covering
a broad time span of 166 years,
the graphs for TB and measles displayed in Fig. 1a,c probably
miss the initial epidemic phase.
\qee{TB1} {\bf Epidemic phase.}\quad
When the prevalence of the infectious agent
(whether bacteria or viruses) in a population 
is fairly low, few people
have been able to develop an immunity. In this phase
the disease manifests itself in the form of periodic epidemic
surges. 
\qee{TB2} {\bf Endemic phase.}\quad
As the outbreaks become more frequent
infection becomes more prevalent with the result that the
disease tends to become endemic. An endemic equilibrium state is
reached when the death rate of the disease has become fairly
stationary. Whereas for rare diseases
such a stationary death rate can be
quite low, in some cases of which TB in the 19th century 
is a good illustration, it can be fairly high.
\qee{TB3} {\bf Fade-out phase through immunity of survivors 
and offspring.}\quad
When the prevalence of a disease is as high as it
was for TB
in the late 19th century, there are more and more survivors 
who develop
an immunity which means a correlative fall in the
number of susceptibles. In this process
the transmission of immunity from mother to child
through breast-feeding played an important role
as was proved by the German biologist Paul Ehrlich
in the early 20th century.\qL
As a result of a decreasing concentration of pathogens in
the environment, the death rate of the
disease starts to fall.  A ``spontaneous'' decrease of this kind
was indeed observed for measles as well as for TB during
several decades prior to the introduction of effective
remedies in the 1950s (see Fig. 1a,c).
\qee{TB4} {\bf Eradication phase.}\quad
With the disease held in check by vaccination and
other remedies and if there is no other carrier which
can constitute a ``safe haven'' for the pathogens involved,
the decrease will go on and will end only
when the disease
is totally eradicated worldwide, a success so far
reached for only a few diseases; back in 1980 smallpox was
the first one.

\qA{Comparative analysis of cancer vs. TB}

Now, let us come to the crucial question of why the
cancer death rate curve is almost flat. Our reasoning
comprises the following steps.
\qee{C1} {\bf Deviant cells are ubiquitous}\quad 
It seems natural to admit that the incidence
of a disease represents an equilibrium between two opposite
forces: the prevalence of infectious agents 
in the environment and the
strength of the immune system with respect to this
specific disease. If there are no pathogens there will
be no disease. As we have seen, this corresponds to
the situation when a disease has been eradicated.
Cancer corresponds to a completely opposite situation
for the analogue of the pathogens are abnormal
cells which grow in a
boundless way. As such deviant cells appear 
continuously in
the body of each and every individual,
cancer resembles a disease brought about by a 
highly widespread pathogen. 
\qee{C2} {\bf Is there a learning curve with regard to the
elimination of abnormal cells?}\quad 
Now, we need to gauge the strength of the
immune system with respect to cancer cells.
Individuals who have recovered from smallpox will
never become ill again which means that their immune
system has been notably strengthened with respect to
this disease. 
Similarly, it was shown by Paul Ehrling that mice
which are fed small but increasing dosages
of a poison (e.g. ricin or abrin) become resistant
to this poison. 
Is there a similar strengthening mechanism
for the immune system against cancer? \qL
It is the task of the killer T-cells to identify and
target abnormal cells. This includes cells 
infected by viruses or bacteria as well as cancer cells.
As the first abnormal cells appear soon after birth
(as seen by the occurrence of some forms of cancer in newborns),
the existence of a learning process would mean that
the immune system will be more and more able
to remove them.
However, the fact that after the age of 10, 
the death rate of cancer starts to increase
and that it continues to go up
steadily with age suggests that no equilibrium is
reached. In a population of 100,000
30-year old persons the immune system will loose its
war against  abnormal cells in approximately
30 cases; at age 40 and 50 that war
will be lost in 60 and 120 cases respectively 
(for data about age-specific death rates see Richmond et al. 2018).
In fact, we do not need to know whether the immune 
system improves or deteriorates with age; what the previous numbers
show is that the imbalance between the rate of apparition of
new abnormal cells and the capacity of the immune system
to remove them increases with age. In other words,
with respect to cancer, the human immune system reveals a structural
weakness. 
\qee{C3} {\bf Cancer in dogs}\quad
In the previous sentence we used the expression
``{\it human} immune system'' which raises the question
of whether this weakness is special to humans or, on the contrary,
is shared by other mammal species. It seems that
the only animals for which the causes of
death are identified with sufficient accuracy are pet dogs and cats.
For both adult dogs and cats cancer is the main cause of death.
Moreover, Fig.1e shows that the trend of the age-specific cancer 
death rate for a group
of 345 Rottweiler dogs is mostly upward.
Although this is a fairly small group
the results give a preliminary indication that the
pattern is roughly similar to what is observed in humans.
In other words, the weakness of the immune systems
is shared by dogs. Probably other mammals would reveal the
same pattern if appropriate data were available.
%
\begin{figure}[htb]
\centerline{\psfig{width=10cm,figure=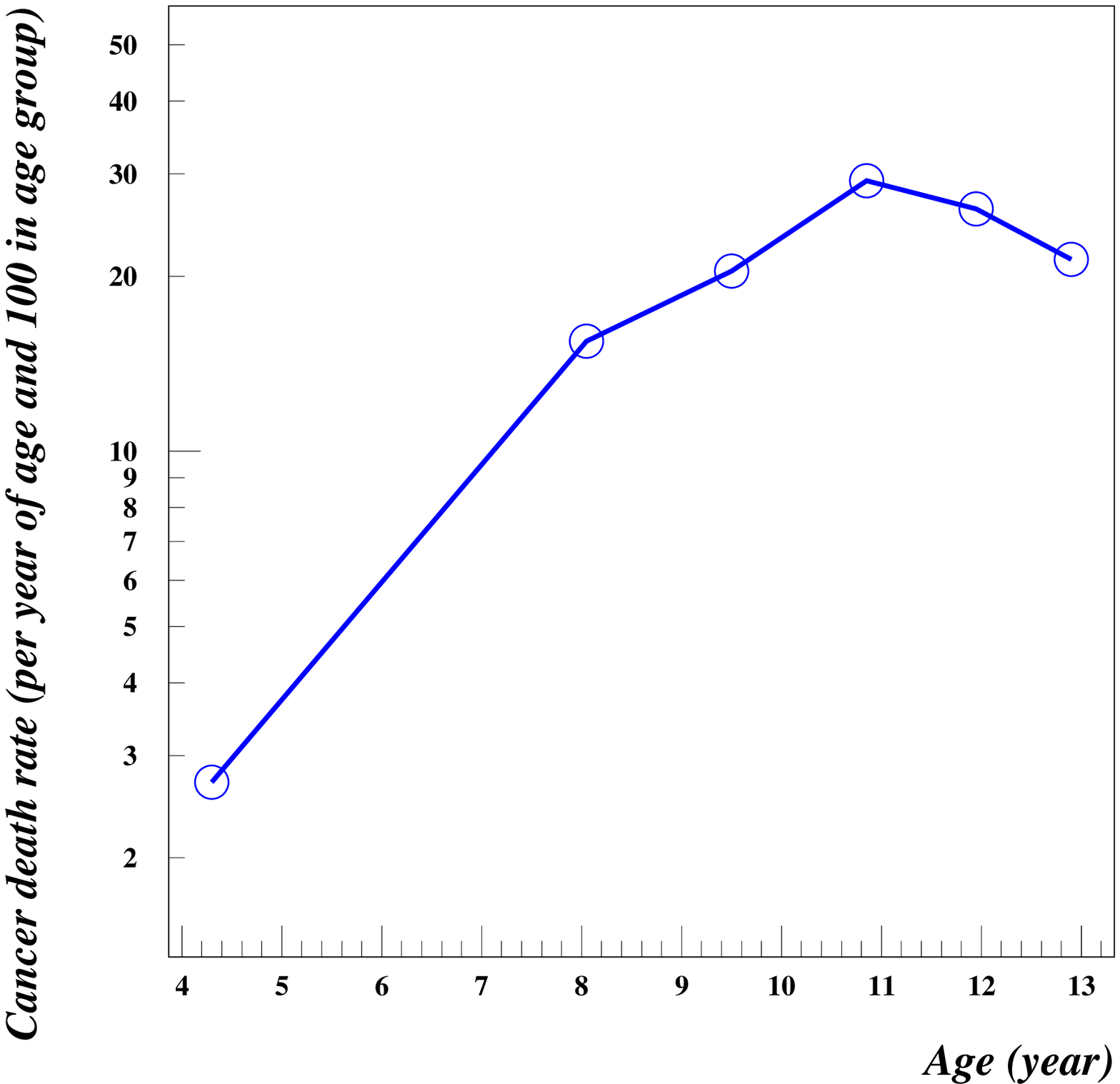}}
\qleg{Fig.\qhu 1e\qhv Cancer death rate for
pet dogs as a function of age.}
{The data are based on information provided by
veterinarians for a cohort of 345 Rottweiler dogs.
With respect to humans the death rate increases
more slowly in old age. However,
in order to asses more reliably the death rate in old
age one would need a larger sample. Here at the age
of 12.5 years there were only 35 dogs still alive.
Between the ages of 4 and 10 there is an exponential
increase described by $ \mu\sim \exp(\alpha t),
\ \alpha=0.43\ \hbox{year}^{-1} $
which corresponds to a doubling time of 1.6 year; this is
almost the same as for humans when renormalized with
respect to the respective life spans, i.e. 1 dog-year equal
some 6 human years.}
{Source: Cooley et al. (2003).}
\end{figure}
%
\qee{C4} {\bf Poor performance of cancer 
immunotherapy and chemotherapy}\quad
For infectious diseases the main 
method of treatment was to strengthen the immune system
through vaccination. Quite understandably, the same
approach was tried for cancer.
Paul Ehrlich (1854--1915),  William Coley (1862--1936)
and others developed cancer immunotherapy%
\qfoot{Interestingly, Ehrlich told his sponsors
that cancer research meant basic
research and that a cure could not be expected soon.}%
. 
They were able to prove that the mechanism of removal of 
abnormal cells by the immune system
is fairly similar to the elimination of pathogens.
However, despite notable progress (for which Ehrlich
was awarded the Nobel prize in 1909), attempted
cures were much less
successful than for smallpox, rabies or other infectious
diseases.\qL 
Ehrlich was not only a pioneer in immunotherapy but
also in chemotherapy. Around 1910 he discovered
the first effective drug treatment for syphilis.
However, when tried for cancer, the approach of
chemotherapy once again proved less effective
than for infectious diseases. 
\qpar

In short, the component of the immune system tasked
with fighting abnormal cells (let us call it 
the {\it endo immune system}) turned out to 
be weaker and more difficult
to strengthen than the component (let us call
it the {\it exo immune system}) destined to fight pathogens.
In a parallel developed below this difference can be seen
as similar to the one between fighting a
foreign war and a guerrilla war.
\qpar

In spite of the differences emphasized in the
previous discussion, the main argument 
of our paper is that better insight 
can be gained by {\it comparing} the 
two diseases. On the one hand, as shown below, they share many
similarities but on the other hand with respect to
treatment they are dramatically different.

\qA{Questions} 

Here are some
of the questions that can be raised.
\qee{1} Both TB and cancer do exist in dormant
mode. In latent TB the persons are infected
with the TB bacteria which however do not multiply; as a result
they  are not ill and not contagious. This state
can last for years. Even without treatment less than
10\% of these persons will develop the disease 
during their lifetime. Similarly, dormant cancer cells
are mutated cells which, however, do not multiply.
As explained below dormancy is a very common occurrence in
many microorganisms. Would it not be of interest to
examine if there are common dormancy mechanisms.
TB and cancer might be a good point where to start such
a comparative study.
\qee{2} Both TB bacteria and cancer cells form clusters.
This is for them a way to be protected against anti-bodies
or drugs. However, the death rate differential suggest
that TB clusters afford much less effective protection.
\qL
In a sense this is a network problem in which the
bacteria or cancer cells are the nodes; such networks
would have the ability to branch out and nucleate to
form secondary networks. Naturally, to define
such networks in a meaningful way would require a
knowledge of key-parameters such as density and
interaction strength.
\qee{3} Whether in TB or in cancer, the creation of new
clusters (i.e. metastasis) is conditioned by the same steps.
(i) First bacteria or cells must be able to leave the
initial cluster. In cancer tumors this means being
able to cross the extra cellular matrix, a layer of 
protein molecules which surrounds the tumor and separates it
from adjoining tissues.
(ii) While on their way the bacteria or cancer cells must
either not be detected or be able to ward off
anti-body strikes.
(iii) Finally, an appropriate niche must be found.
Our subsequent analysis suggests that TB myco bacteria
and cancer cells have the same preferences, namely
the lungs and the other organs that we examine but that
they both avoid skeletal muscles.\qL
Whether or not we have a good understanding of these
steps can be tested through our ability to explain the
differential outcomes.
\qpar

In the next section additional explanations
are provided about the mechanisms involved in the
previous points.

\qI{Similarities between TB and cancer}

\qA{Treatment}

The impressive difference suggested by Fig 1a,b
is an enigma for
in a sense the situations of the two
diseases  with respect to treatment appear fairly similar.
\qbu Drugs are available which kill the agents
responsible of the disease, {\it Mycobacterium
tuberculosis} on one hand and cancer cells on the other
hand. The mechanism of action of antibiotics and
drugs used in chemotherapy are very much the same in the
sense that both drugs are only effective when the bacteria or
cells are in division stage. 
\qbu There are two main mechanisms 
through which TB bacteria and cancer cells
can protect themselves. One is to form tight clusters that anti-bodies
cannot penetrate. The second is to form 
dormant cells which cease dividing but survive in a quiescent state
while waiting for appropriate environmental conditions.
Among microorganisms this a very common
mechanism. For instance, in {\it C. elegans} it is called
the {\it dauer} stage, in the rotifer {\it Brachionus plicatilis}
the so-called diapausing eggs may remain in an arrested stage
for long periods until conditions become more favorable
(Garcia-Roger et al. 2005).
\qpar

In other words, TB and cancer treatments are confronted
to similar problems: how to break into clusters and
how to get rid of the dormant cells.
\qpar

It may be tempting to say that in these two tasks antibiotics are
doing a better job than cancer chemotherapy drugs. However,
this would ignore the fact that 
between 1850 and 1950 (beginning of antibiotics treatment)
TB mortality rate had fallen from 350 to 40 per 100,000.

\qA{Worsening environmental factors}

If antibiotics were only of marginal importance, what were
the other factors? The BCG (Bacille Calmette Gu\'erin)
vaccine was first used on humans in 1921 in France 
but it was not until after World War II that it received widespread
acceptance elsewhere. As it arrived at the same
time as the antibiotics it can hardly be credited with the
sharp reduction since 1850. So, what were the 
factors at work in the 100 years until 1950?
Mainly environmental factors and better living conditions.
More specifically, this meant preventing contagion and
allowing the organisms of patients to fight the
disease in the best conditions possible (sanatorium,
rest, healthy apartments).
\qpar

Why did progress of a similar kind not take place for cancer?
This is an interesting but open question.
However, it can safely be said that with respect to
cancer, environmental conditions rather
deteriorated. 
One may mention
food additives, pesticides used in agriculture or X-rays.
Except the last one,
most of these factors are not easy to
quantify. 
An introductory discussion of the X-ray issue can be
found in Roehner (2002, p.42-45).
\qpar

At first sight it may seem possible to estimate the
effect of environmental factors through the incidence rate.
However, pre-death medical diagnosis is very dependent upon
available identification means. For instance, between 1982 and 1989
the incidence rate of breast tumors smaller than 1cm 
increased from 10  to 40 per 100,000 women whereas in the same time
interval, the incidence of tumors of more than 3cm remained
unchanged at a level of 35 per 100,000 (Garfinkel et al. 1993).

\qA{Granulomas versus tumors}

The mechanism of TB propagation in an organism
can be closely investigated in experimental TB induced
in animal models. Zebrafish ({\it Danio rerio}) appear
as a particularly convenient model (Leeuwen et al. 2015).
The development of granulomas are a key feature of human
TB and zebrafish studies provided a dynamic insight into
what, hitherto, was largely seen as static structures.
Granulomas are preferentially formed in fatty tissue
and they are the first step through which infection
can be clearly identified. 
In short, they play the same role as
primary tumors in cancer. 
Moreover, observations on zebrafish 
showed that infected macrophages can detach from
established granulomas and wander off to other
organs to form secondary granulomas. Clearly,
such a process parallels the dissemination of
cancer cells which leads to secondary tumors.

\qA{Differences that are more apparent than real}

At first sight, tuberculosis (TB) and cancer appear as very
different diseases. Whereas TB is due to the
proliferation of a bacteria, {\it Mycobacterium tuberculosis} (MT),
cancer results from the boundless reproduction of 
body cells which have freed themselves from
the controls generated by their neighbors. TB
starts with a foreign pathogen whereas in cancer it is a body
cell which spins out of control.
\qpar

However, if we look at it more closely, the mechanisms are not
too different for indeed through a kind of reprogramming process
pathogens are able to twist the behavior of
host cells (in the case of viruses) or macrophages
(in the case of bacteria) 
which try to eliminate them. For instance, 
once a bacterium has been ``imprisoned'' into
a vesicle of a macrophage%
\qfoot{The diameter of a human macrophage is about 20 micrometers
which is 10 times more than the length of MT bacteria.}
,
a common trick (used for instance
by MT or {\it Legionella pneumophila}) is to prevent the
injection into the vesicle of enzymes which would
kill the pathogen. In short,
by manipulating their host successful pathogens are
able to use it as a vehicle for proliferation.
\qpar

Conversely it was estimated (Greenwood 2012)
that about 16\% of worldwide cancers
are triggered by viruses or bacteria: for stomach cancer it is
Helicobacter pylori, for lymphoma it is the Epstein-Barr virus,
for cervical cancer it is the papilloma virus,
for liver cancer it is hepatitis B or C.
\qpar

There are also similarities in subsequent stages.
\qbu One knows that the formation
of a tumor is a multistage process; so is also
the development of TB as illustrated by the following features.
(i) In only 10\% of infected persons
will TB manifest itself. The other 90\% will remain
asymptomatic. 
(ii) As already mentioned as TB develops, 
one sees the formation
of granulomas which are nodules of a diameter of about 3mm
in which
pathogens and neutralized macrophages are clustered together.
\qbu Primary cancer tumors can develop in many organs. 
Although
pulmonary TB is the most frequent, TB can also develop in
several parts of the body: bones, skin, intestines, meningitis,
genito-urinary organs. On the other hand there are some locations
(e.g. skeletal muscles)
where the development of both cancer and TB is uncommon.
\qbu Although fairly rare, TB of the breast (tuberculosis
mastitis) does occur and it is
far more frequent (about 20 times)
in women than in men (Wilson et al. 1990); it is particularly likely
in reproductive age.
As a matter of fact, being irregular and hard,
TB breast lesions
may be indistinguishable from breast carcinoma 
(Baharoon 2008).
\qbu One of the most important properties of cancer cells is their
ability to spread to other organs where 
they may form secondary tumors. There is a parallel
in TB in the sense that the granulomas can migrate to
other organs such as the liver or the kidneys. 
In developed form such cases are
called miliary TB; they represent 20\% of the extra-pulmonary cases. 
\qpar

Although too short and schematic, the previous description
suggests that it is not completely absurd to think that
there may be connections between the development of TB and cancer. 

\qI{Purpose and method}

\qA{Purpose}

The objective of this paper is very simple.\qL
If, as argued above, there is indeed a connection between
the growth of TB and cancer in various organs, can it be 
that these links manifest themselves by similar responses
of various organs? How can one design an operational
procedure that will unravel such possible links?
\qpar

It turns out that among the set of persons who
will develop a cancer during their lifetime
the probability $ C_b(t) $ that this cancer will 
affect the brain
is highest around the age of 10 years: $ C_b(10)=23\% $ (Fig. 1).
At older ages this probability falls off sharply.
At the age of 25 it is down to 5\%, and at the age of 50 to
$ 1\% $. Do we currently have a clear understanding
of why $ C_b(t) $ is highest in childhood? To our
best knowledge, it does not seem so.
Now, as can be seen in Fig. 1, the probability $ T_b(t) $ for
developing TB affecting the brain is also maximum at the age
of 10 and falls off in older ages even more sharply than
$ C_b(t) $. Subsequently, for the sake of brevity,
a curve such as $ C_b(t) $ will be called the {\it age-profile}
of brain cancer.
\qpar

It could well be that the similarity in the age-profiles
$ C_b(t) $ and $ T_b(t) $ 
is merely a coincidence. However, if for a number of other
organs the same similarity is observed, it may suggest
that some organ-specific factors are at work which account for
similar responses for the two diseases. 
\qpar

What can be the practical interest of these observations?
Taken alone, they will not give us a full fledged explanation
of why such brain diseases appear at that specific age.
However, they suggest that the idiosyncrasies of the
organ under consideration
plays a more important role than the nature
of the disease itself. For instance, as far as brain
diseases are
concerned, the screening defects of the blood-brain
barrier may be the key%
\qfoot{Some of the ``holes'' of this barrier are
described in Kim (2010 p.33-34) and Sorge et al. (2012).
For instance, the barrier can be penetrated by
using so-called ``Trojan horse'' mechanisms.}%
.
In other words, in this shield versus sword issue,
our observation
will tend to focus attention on the macrobiological properties
of the shield.

\qA{Method}

In the previous subsection we defined our objective and
we broadly delineated how the observation will be organized.
Presently, we will explain how in practice it can be implemented.
\qpar

The main difficulty is the following. \qL
As we wish to
explore the affinity a disease ($ \alpha $) has for a 
specified organ ($ S  $)
we would need incidence data rather than mortality
data. As a matter of fact,
mortality data mix two very different
effects (i) the prevalence of $ \alpha $ in $ S $ 
(ii) how effectively this specific disease can be cured. 
As an extreme case, for a disease 
whose cure rate is 100\% or which simply is not a deadly 
disease (as for instance is the case of most
forms of herpes)
it is altogether impossible to get any information
about incidence from mortality data.
\qpar
Because nowadays in all developed countries TB is cured
in very effective ways%
\qfoot{In 2015 in the US the total number of TB deaths 
was 470 which is 152 times less than in 1934.
In 1910 there were even 2.4 times more TB deaths than
in 1934.
For cancer it is a different situation in the sense
that there were 140,771 deaths in 1934 and 612,207 in
2015.}%
, 
it is impossible to use present-time
data for our investigation.
On the other hand, if we consider US data from the early
20th century there will also be two difficulties.
(i) The international classification of diseases
will be very different from the present-day classification
and it will be much less detailed. Thus for TB
there will be only 5 types and they will include labels such
as ``Pott's disease'' (TB of the spine) or ``White swellings''
(TB of the joints) with which we are no longer familiar.
(ii) In 1910, only 58\% of the US population was included
in the federal death registration area.  This would reduce
the number of deaths by a factor of two.
Starting from 1933, the death registration area included
100\% of the population. This led us to 
the methodological choice of using data from the period 1934-1945.
\qpar
In 1934 there were substantial numbers of deaths in almost
all types of TB or cancer except for a few.
We have been using averages over a number of successive years
in order to reduce the statistical fluctuations.
Naturally, the fluctuations are conditioned by the
number of deaths. As an illustration, one can mention the case
of skin TB for which there were less than 10 deaths in almost
all age groups except in the oldest ones. In this case
the smoothing process required to take averages over 10
successive years that is to say from 1934 to 1943.

\qI{Results}

If one can assume that the death records have been filed
correctly, all disease locations documented in US mortality
data are for {\it primary} lesions.
The comparison between the profiles of TB types
and cancer types is shown in Fig. 2a,b.

%
\begin{figure}[htb]
\centerline{\psfig{width=16cm,figure=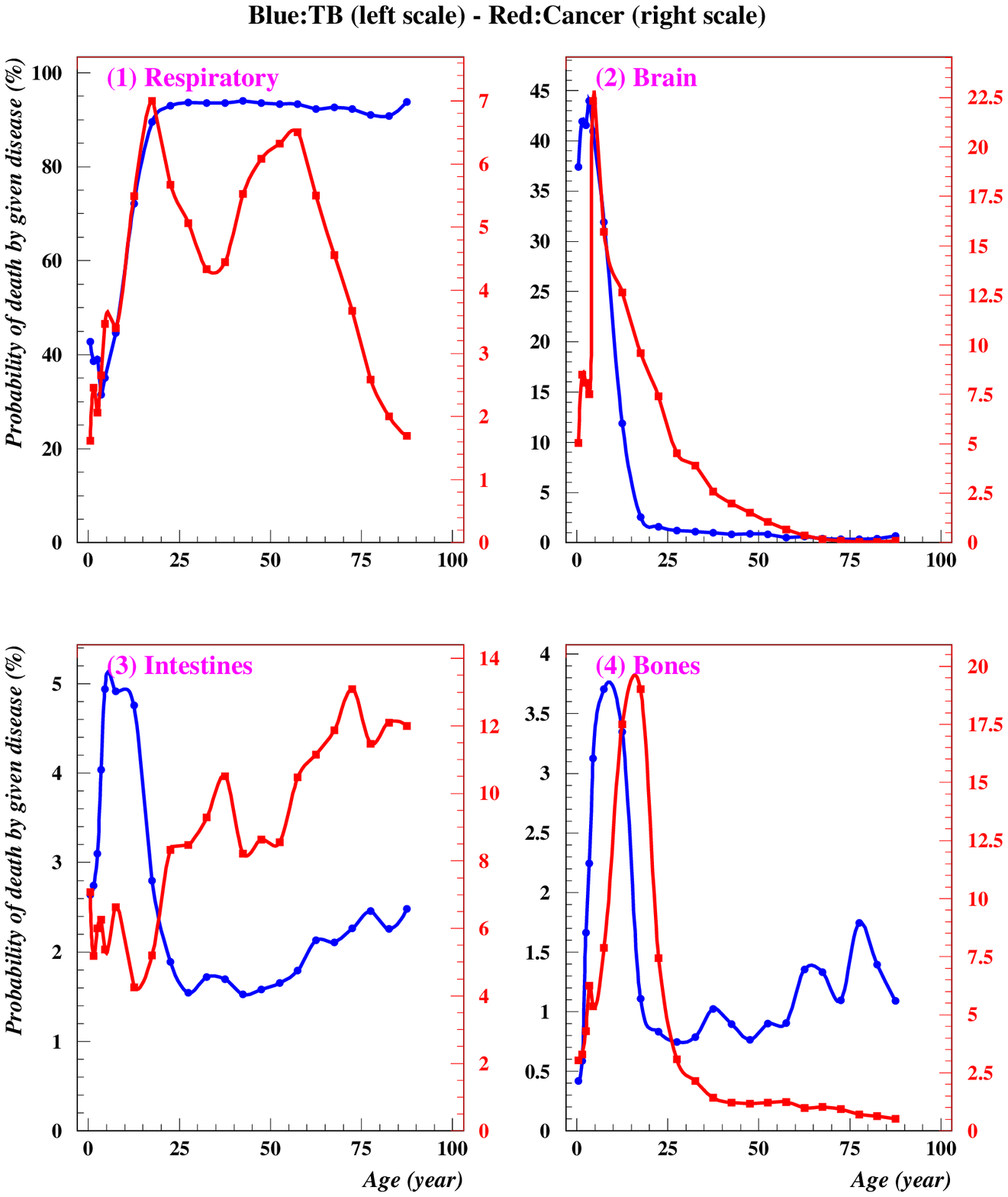}}
\qleg{Fig.\qhu 2a\qhv (continued in Fig. 1b). 
Comparison between the conditional probabilities
represented by the ratios:
(deaths by TB type)/(all TB deaths) (blue line with round dots) 
and
(deaths by cancer type)/(all cancer deaths) (red line with squares),
USA 1934-1943.}
{The left-hand scale refers to TB whereas the right-hand 
scale refers to cancer.}
{Sources: Bureau of the Census: Mortality Statistics 1934-1936,
and  Bureau of the Census: Vital Statistics of the United States,
Part 1, 1937-1943.}
\end{figure}
%
\begin{figure}[htb]
\centerline{\psfig{width=16cm,figure=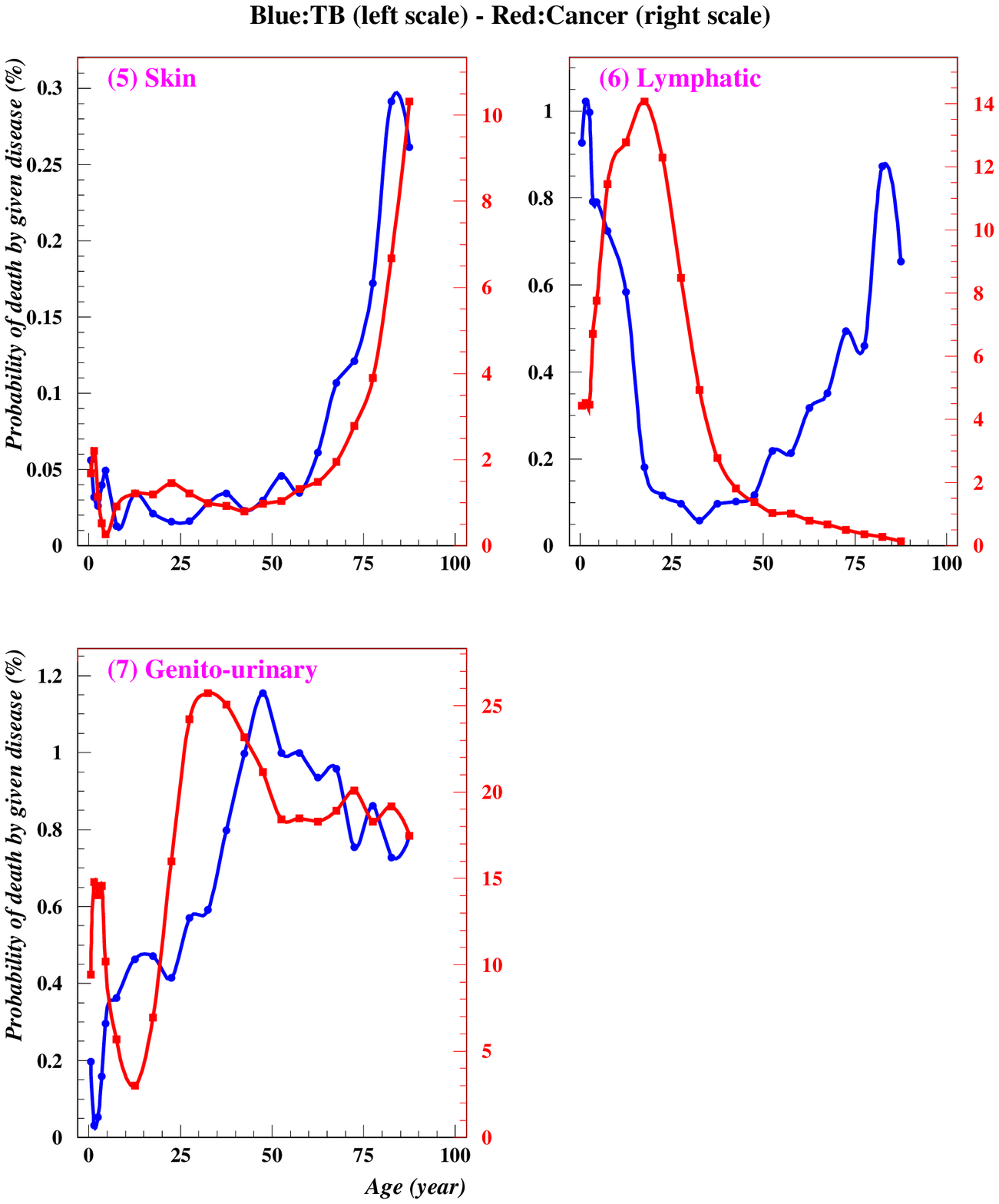}}
\qleg{Fig.\qhu 2b (continued from Fig. 1a).\qhv 
Comparison between the conditional probabilities represented by
the ratios:
(deaths by TB type)/(all TB deaths) (blue line with round dots)
and
(deaths by cancer type)/(all cancer deaths) (red line with squares), 
USA 1934-1943.}
{The left-hand scale refers to TB whereas the right-hand 
scale refers to cancer.
The TB ratios represent the conditional probability of
dying at a given age from a specified type of TB within
the set of all people dying from TB. There is
a similar definition for cancer.
The 7 cases shown in Fig. 1a,b do not result
from a selection; they represent {\it all cases}
for which data are available. The correlations, 
confidence intervals (for a probability level of 0.95)
and ICD codes are given below in the 
following form: cor, CI, {\color{blue} TB}/{\color{red} C}. 
(1) cor $ = 0.55\ (0.17,0.79) $, {\color{blue} 23}/{\color{red} 47};
(2) cor $ =0.67\ (0.35,0.85) $, {\color{blue} 24}/{\color{red} subclass of 53};
(3) cor $ =-0.560\ (-0.81,-0.24) $, {\color{blue} 25}/{\color{red}
  subclass of 46};
(4) cor $=0.46\ (0.046,0.74) $, {\color{blue} 26+27a,b}/{\color{red} 
    subclass of 53};
(5) cor $ =0.91\ (0.79,0.96) $, {\color{blue} 28}/{\color{red} 52};
(6) cor $ =0.01\ (-0.41,0.43)$, {\color{blue} 29}/{\color{red} 72b};
(7) cor $ =0.56\ (0.18,0.79)$, {\color{blue} 30}/{\color{red} 46+51}.
All correlations are significant 
except (3) and (6).
Moreover under an age shift of 10 years the  two curves of (6)
become correlated too.}
{Sources: Bureau of the Census: Mortality Statistics 1934-1936,
and  Bureau of the Census: Vital Statistics of the United States,
Part 1, 1937-1943.}
\end{figure}

As a matter of curiosity,
one may wish to know to what extent the cancer profiles of 1934 
differed from those of 2015.
Of the 7 profiles, only two were markedly different, namely
(i) cancer of the respiratory system and (ii) skin cancer.
For (i) in 2015 there was only one peak near the age of 65
The narrow peak centered on the age of 20 does no longer
exist. For (ii) it so to say the opposite in the sense
that in 2015 in addition to the ratio which surges up beyond
the age of 75
there is also around the age of 25 a major peak which reaches
the level of 4\%. 
\qpar

The TB-cancer correlations are given in the caption of
Fig. 2. They show that in 5 of the 7 cases (i.e. 71\%)
there is a significant correlation. For these 5 cases
the average correlation is: $ r=0.61\pm 0.11 $.
\qpar
In addition two points should be observed.
\qbu Even for the ``intestines'' and ``lymphatic'' cases
for which there is no correlation the profiles share
some common features. Thus, for ``intestines'' after the age of
25 the two ratios do not fall to zero (as is the case for ``brain''
or ``bones'') but increase more or less steadily 
(applying a smoothing moving average would almost eliminate
the ups and downs of the cancer profile). Moreover,
an age shift of only 10 years would bring the peaks of
the two curves of the lymphatic case together.
\qbu It must be realized that the profiles are of very
different shapes which makes similarities arising merely
by chance fairly unlikely.

\qI{A conjecture to explain the treatment conundrum}

\qA{Death rate levels versus death profiles}

What is the present-day situation regarding TB and cancer?
As shown in Fig. 3, roughly speaking in terms of
order of magnitude TB and cancer death rates differ
by a factor 100 at all ages. However,
contrary to their levels, the age-profiles of death rates 
are  not very different.
The parameters of the two
laws differ by 20\% on average (23\% for the
exponent $ \gamma $ and 18\% for the doubling time $ T $.)
%
\begin{figure}[htb]
\centerline{\psfig{width=14cm,figure=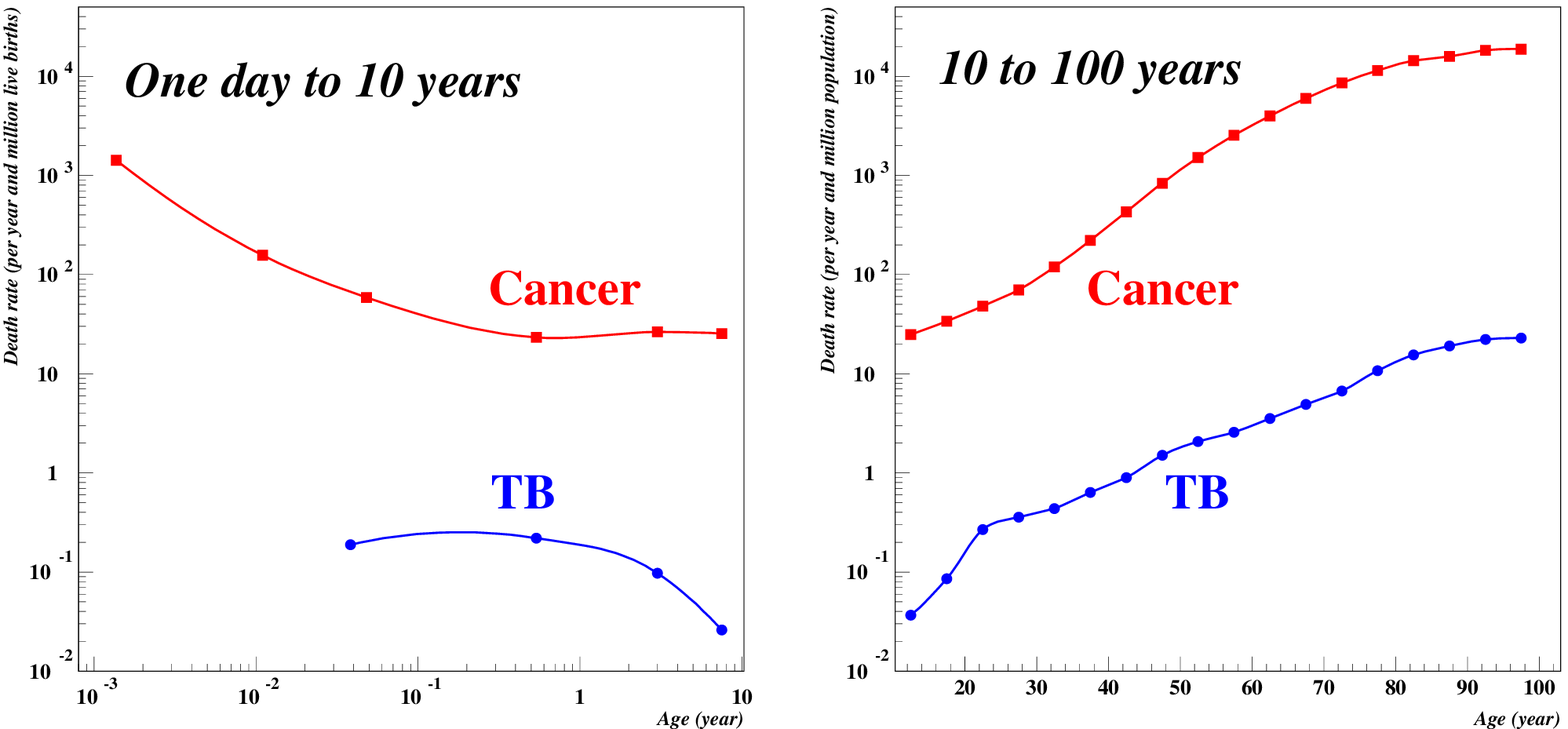}}
\qleg{Fig.\qhu 3\qhv Comparison between the death rates
of TB and cancer, USA 1999-2015. Left: age from 1 day to 10 years,
right: age over 10 years.}
{Whereas the death levels are very different, the shapes
of the age-specific death rates are fairly similar.
For all causes of death, in the 0-10 year age range
the death rate is an hyperbolic function of age of the form: 
$ \mu_b\sim 1/t^{\gamma} $ (Berrut et al. 2016).
For TB: $ \gamma=0.33\pm 0.3 $,
for cancer: $ \gamma=0.43\pm 0.2 $. In the range 10-100 year,
it is the Gompertz law which applies, that is to say an
exponential increase: $ \mu \sim \exp(t/\tau) $.
The doubling times $ T=\tau \log(2) $ expressed in years 
are as follows. TB: $ T=9.7\pm 0.8 $, cancer: $ T=7.9\pm 0.7 $.}
{Source: CDC Wonder database 1999-2015 (underlying causes of death).
As of April 2017 the relevant web-address was:
https://wonder.cdc.gov}
\end{figure}

\qA{Birth defects versus wear-out}

The curves of
Fig. 1 suggest that there is a connection in the way
TB and cancer lesions develop in different organs. 
A rough distinction can be made between the 
lesions (e.g. 2,4,6) which develop mostly in young age
(and are likely due to birth defects) and the 
lesions (e.g. 1,3,5,7) which develop mostly in old age (and are 
likely due to a wear-out process).
\qpar

Such conclusions would have been impossible to reach by
using death rates instead of death ratios. Indeed,
as shown in Fig. 2, 
age-specific death rate curves in the adult age-range
are increasing exponential
functions, a feature that
would preclude any shape comparison.
On the contrary, the age-profiles of the death ratios
have very contrasted shapes
that can be compared significantly. The aim of this article was
to convey our belief that the age-profiles offer
useful ``signatures'' of the corresponding diseases. 

\qA{Conjecture for the treatment conundrum}

Despite the similarities that we were able to pinpoint in this
article, TB and cancer are extremely different in
terms of treatment effectiveness. Fig. 1a,b shows that in the past 100
years TB mortality was divided by 1,000 whereas cancer mortality
remained basically the same. This is what we call
the {\it treatment conundrum}. How can it be explained?
In the following subsection we propose a testable conjecture.

\qA{Army vs. police}

In any sovereign country a clear distinction is made between
army and police. The role of the army is to ward off
foreign attackers and to destroy the forces who were able to
penetrate on the national territory. In contrast the action
of the police is directed against domestic elements who
are sources of disorder or threaten the nation's security.
Whereas waging foreign wars is a well defined concept,
fighting domestic enemies of the state is a more murky task.
The two notions tend to overlap in civil wars
or guerrilla warfare. The specificity of guerrilla warfare is
that the insurgents hide among the population and are
therefore not easy to identify and destroy. 
There are no definitive victories. Once a region has been cleaned
fighting may resume elsewhere.
\qpar

A similar distinction can be made for the immune system.
Bacteria, viruses, fungi, parasites are foreign invaders
which will be identified, targeted and destroyed by
the lymphocytes (B and T cells), antibody proteins and
macrophages. On the contrary, cancer cells are domestic
enemies. For the sake of brevity, the part of the
immune system which fights foreign invaders will be
called the {\it exosystem} whereas the part directed against
domestic deviants will be called the {\it endosystem}%
\qfoot
{It would seem that the exosystem has received more attention 
than the endosystem. On the website of the Nobel Committee
there is a fairly detailed presentation of the immune system
destined to the general public. In the 5 pages there is only
one line which refers to cancer cells. We are told
that the killer T cells (one of the two kinds of T cells)
are specialized in attacking {\it cells} of the body infected
by viruses or bacteria as well as cancer cells.}%
.
\qpar

It may be useful to note that in fact the exo- and endosystems
overlap. A clear confirmation is the fact that the BCG vaccine
against tuberculosis is a remarkable booster of the immune
response to bladder cancer. 
Since the late 1970s it is known that
instillation of BCG into the bladder brings about 
a local immune reaction against the
tumor. Moreover recurrence is prevented in up to 67\% of
cases (Wikipedia article entitled ``BCG vaccine'').

\qA{A testable explanation of the treatment conundrum}

The distinction between the exo- and endosystem
suggests a possible  explanation.
\qpar

We have observed that partisan fighters are much more difficult
to identify than foreign armed forces. It is natural
therefore to make the same assumption for the immune system.
More precisely, we make the following assumption.
\qdec{{\it \color{blue} Exo-endo conjecture} \quad 
For an organism which
enjoys optimal living conditions, at any age the 
exo immune system is more effective than the endosystem
in the sense that the natural mortality rate is
much higher for endodiseases (like cancer)
than for exodiseases (like tuberculosis).}
\qpar

This conjecture explains that the present-day death rate
is 1,000 times lower for TB than for cancer but it does
not explain why the two rates were almost the same in the
early 20th century. This feature can be explained by 
population movements (particularly rural
flight%
\qfoot{In all industrialized countries
there were large-scale population migrations
from the countryside to cities and their factories.}%
)
 followed by a selection process. 
What do we mean more precisely?

\qA{Role of rural flight}

Unfortunately we do not know early  prevalence levels
around 1800 respectively in rural and urban places
but irrespective of where it was higher the contacts
between rural and urban people must have increased
the number of cases especially in the cities because
of a higher population density (with respect
to the influence of density see Li et al. 2018)
and poorer living 
conditions. This resulted in a great TB epidemic
observed in all European countries.
\qpar

As in all epidemics the persons whose
immune system was able to protect them survived
along with their children. We are not used to
view TB as an epidemic but one should remember
that from infection to death or recovery
the time constant of this disease is of the order
of several years. For influenza it is a few weeks
that is to say some 50 times less. Whereas an influenza
epidemic lasts half a year, a TB epidemic may
last 25 years. 
\qpar

The fall of the TB death rate between 1850 and 1950
is often explained by better living conditions.
While living conditions may have played a role,
in our conjecture the bulk of the decrease was due
to the selection process through which people
whose immune system could not fight the disease
were eliminated. The people who survived were those
who had a natural immunity or whose immune system
was able to adapt.\qL
If this explanation is correct one should observe
a correlation between the development of population
interaction (particularly the
rate of rural flight) and the death rate due to TB.
\qpar

There remains a question however.\qL
Should one not observe a similar selection process
with respect to cancer?\qL
For such a selection to operate, the persons
who have weak immunity should not be able to transmit it
to their children.\qL
This was indeed the case for TB whose peak rate
occurred in the 20-30 age group (see Richmond et al. 2018)
but is not at all the case for cancer whose death
rate is very low for the 20-30 age group
and becomes substantial only after the age of 60.

\qI{Conclusion}

\qA{Similarities and differences between TB and cancer}

The main message of this paper was to emphasize the
similarities between TB and cancer. However, this
left open the question of why there was a tremendous
difference in the outcome of treatment.
To explain it we proposed what we called the exo-endo
conjecture. How can it be tested?
\qpar

The simplest test would be a comparison of mortality
rates of exodiseases and endodiseases in other mammals.
The Wikipedia article entitled ``Cancer in dogs'' gives the
following information.
\qdec{Dogs can develop a variety of cancers and most are very similar
  to those found in humans. 
Cancer is the leading cause of death in dogs.
It is estimated that 30\% domestic dogs will develop cancer, 
which is the same incidence as in humans.}
\qpar

A more thorough investigation of the respective frequency 
of exo- and endo-diseases in mammals will be postponed 
to a subsequent paper.

\qA{Extension of the present study}

 For the purpose of this paper we had by necessity to use
data from the first half of the 20th century because
our main requirement was to have enough death cases
to keep the statistical fluctuations under control.
How can we extend this analysis to a greater number of lesions?
\qpar

If, instead of using death rates we could use incidence rates
the number of cases would be multiplied by a large factor.
This factor depends of course upon how lethal
a lesion is; for instance lesions of the thyroid can
be cured with high probability which means that the 
incidence rate may be 20 or 30 times higher than the 
death rate. This will allow an interesting extension.
\qpar

TB and cancer are not the only diseases which can 
target different organs; syphilis or diabetes are other
examples. This will provide a second extension.
It will be interesting to see if there are meaningful
connections between such families of diseases.

\appendix

\qI{Appendix A. Importance of thinking by analogy}

Readers may wonder what can be gained and learned by 
emphasizing analogies between TB and cancer.
Perhaps the best answer is to observe that
throughout the history of science analogies
have permitted great progress. This is explained
in Appendix A.
\qpar

It is not always realized that
throughout the development of physics
from the 16th century up to nowadays
physicists have relied on thinking by analogy.
The purpose was always the same, namely to
establish a bridge between phenomena which were
fairly well understood and others which were still
mysterious. Seen in this perspective, the role
of thinking by analogy should not be limited
to physics. In any field, when facing uncharted
territory, one needs some guidance. While assistance
may come in many different forms, analogy with
similar but simpler cases is certainly one of them.
As an example
Mendel's experiments provided the framework
on which the whole field of genetics could develop.

\qA{Rigid versus open scientific thinking}

At first sight, thinking by analogy may
not appear very scientific. It is true that 
it cannot
prove itself to be correct but at least
analogy reasoning
can lead to fruitful conjectures. 
It may well be that in their concern to appear
highly ``scientific'', biologists 
refrain from relying on analogy arguments.
\qL
The very structure of present-day biological
papers, namely: 
(i) Problem 
(ii) Material and method 
(iii) Results and discussion, 
may be appropriate for well defined issues but not
for opening new vistas in unexplored territory.
In physics such a rigid framework would be 
suitable for engineering
problems but certainly not for developing the
new concepts and ground-breaking ideas of a new field like
quantum mechanics. New visions require imagination
and creativity and most often such advances are 
channeled and guided by analogy thinking. 
\qpar
The purpose of the present appendix is to
make biologists realize that even in physics, the most
successful among scientific fields, analogy has
played a great role at some crucial junctures.
In the hope that they may convince doubtful readers, 
we give below
a number of illustrations.

\qA{Examples of progress brought about by analogy thinking}

In the late 16th and early 17th century the question
of how two masses can attract one another without apparently
any contact was a key issue for astronomy as well as for the
question of free fall. From Galileo to Kepler to Newton
electrostatic and magnetic attraction provided welcome
examples supporting the conjecture of the
existence of a gravitational force.
Nowadays, the mechanism of gravitational
attraction explains many phenomena, from the fall of an apple,
to the ``fall'' of the Moon toward the Earth, to the formation
of black holes, to the rotation of spiral galaxies.
\qpar

Three centuries later, in the discussions which led to
the creation of quantum mechanics, the analogy with
optics, namely geometrical optics on the one hand and physical
optics on the other hand, 
played a great role for it facilitated the adoption of
wave-particle duality.
\qpar

Fifty years later, in an attempt to
describe strongly interacting particles, physicists
created what is now called quantum chromodynamics (QCD)
and for that purpose, as a blue print, they used
quantum electrodynamics (QED), that is to say
the study of particles which have only electromagnetic
interactions.
\qpar

Some two decades later, Yang-Mills
gauge theories took model on gauge invariance
in electromagnetism%
\qfoot{That is to say
the fact that there is a degree of arbitrariness 
in the definition of the  scalar and vector potentials.},
see Yang et al. (1954).
\qpar

The role of analogy in the history of physics was 
studied and emphasized by the French physicist
Pierre Duhem as attested
by the following excerpt (Duhem 1906, p. 140).
 ``The history of physics shows
that the search of analogies between different categories
of phenomena may have been the most productive of
approaches tried by theoretical physics''.
\qpar

It should not come as a surprise, therefore, that 
when they turn to the fields of biology and medicine, physicists
adopt a comparative perspective with the purpose of
finding parallels and common rules for apparently
unrelated phenomena.
\qpar

It is hoped that the macro-biological perspective 
that we develop in the present paper may prove
of value in complement to the highly detailed
descriptions permitted by the techniques
of molecular biology and genetic sequencing.

\vskip 10mm

{\bf References}

\qparr
Baharoon (S.) 2008: Tuberculosis of the breast.
Annals of Thoracic Medicine 3,3,110-114.

\qparr
Berrut (S.), Pouillard (V.), Richmond (P.), Roehner (B.M.) 2016:
Deciphering infant mortality. 
Physica A 463,400-426.\qL
[The initial version of this paper is available on the
arXiv website at the following address:
https://arxiv.org/abs/1603.04007]

\qparr
Berrut (S.), Richmond (P.), Roehner (B.M.) 2017:
Age spectrometry of infant death rates as a probe of
immunity: identification of two peaks due to viral
and bacterial diseases respectively. 
Physica A 486,915-924.

\qparr
Berrut (S.), Richmond (P.), Roehner (B.M.) 2018:
Excess tuberculosis mortality in young women: high accuracy
exploration.
Physica A (to appear).

\qparr
Bureau of the Census: Mortality Statistics (various years until 1936).
Washington, Government Printing Office.\qL
[In 1937 the name of this periodical was changed into:
``Vital Statistics of the United States, Part I'' as indicated in the
following reference.] 

\qparr
Bureau of the Census: Vital Statistics of the United States (various
years starting in 1937). Part 1: Natality and mortality data for
the United States tabulated by place of occurrence with supplemental
tables for Hawaii, Puerto Rico, and the Virgin Islands.
Washington, Government Printing Office.\qL
[The volumes of both the ``Mortality Statistics'' and ``Vital Statistics''
are available online. As of April 2017 the relevant address was:\qL
https://www.cdc.gov/nchs/products/vsus.htm\#1950]

\qparr
Cooley (D.M.), Schlittler (D.L.), Glickman (L.T.), Hayek (M.), 
Waters (D.J.) 2003: Exceptional longevity in pet dogs
is accompanied by cancer resistance and delayed onset of
major diseases.
Journal of Gerontology, Series A: Biological Sciences and
Medical Sciences, Vol. 58A,12,1078--1084.

\qparr
Dales (L.G.), Kizer (K.W.), Rutherford (G.W.), Pertowski (C.A.),
Waterman (S.H.), Woodford (G.) 1993:
Measles epidemic from failure to immunize.
Western Journal of Medicine 159,455-464.

\qparr
Duhem (P.) 1906: La th\'eorie physique: son objet, sa structure.
[The structure and purpose of physical theories.]
Chevalier et Rivi\`ere, Paris.

\qparr
Garfinkel (L.), Boring (C.C.), Heath (C.W.) 1993:
Changing trends. An overview of breast cancer incidence and
mortality.
National Conference on Breast Cancer, Boston, 26-28 August 1993.

\qparr
Garc\'{\i}a-Roger (E.M.), Mart\'{\i}nez (A.), Serra (M.) 2005:
Starvation tolerance of rotifers produced from parthenogenetic eggs
and from diapausing eggs: a life table approach.
Journal of Plankton Research 28,3,257-265.

\qparr
Greenwood (V.) 2012: 16\% of cancers are caused by viruses or
bacteria. Discover Magazine 9 May 2012.

\qparr
Grove (R.D.), Hetzel (A.M.) 1968: Vital statistics rates in the
United States 1940--1960. United States Department of Health,
Washington. 

\qparr
Kim (K.S.) 2010: Acute bacterial meningitis in infants and children. 
Lancet Infectious Diseases 10,1,32-42.

\qparr
Koch (R.) 1884: Die \"Atiologie der Tuberkulose. 
Mitteilungen aus dem Kaiserliche Gesundheitsamte 2,1-88.

\qparr
Leeuwen (L.M. van), Sar (A.M. van der), Bitter (W.) 2015:
Animal models of tuberculosis: zebrafish.
Cold Spring Harbor Perspective in Medicine 5,1-13.

\qparr
Li (R.), Richmond (P.), Roehner (B.M.) 2018: Effect
of population density on epidemics.
Physica A (to appear).

\qparr
Linder (F.E.), Grove (R.D.) 1947: Vital statistics rates in the
United States 1900--1940. United States Public Health Service,
Washington.

\qparr
Massachusetts deaths 2007: Report by the Bureau of Health
Information Statistics. Massachusetts Department of 
Public Health, April 2009.

\qparr
Richmond (P.), Roehner (B.M.) 2018:
A 2-d classification of diseases based on age-specific death
rates.
Physica A 492,2281-2291

\qparr
Roehner (B.M.) 2002: Patterns of speculation. A study
in observational econophysics. Cambridge University Press,
Cambridge (UK).

\qparr
Sorge (N.M.), Doran (K.S.) 2012:
Defense at the border: the blood–brain barrier versus bacterial
foreigners.
Future Microbiology 7,3,383-394.

\qparr
Walt (M. van der), Lancaster (J.), Shean (K.) 2016:
Tuberculosis case fatality and other causes of death among
multidrug-resistant tuberculosis patients in a high HIV prevalence
setting, 2000-2008, South Africa.
PLoS One, 11,3, 7 March 2016.

\qparr
Wilson (J.P.), Chapman (S.W.) 1990: Tuberculous mastitis.
Chest 98,1505-1509.

\qparr
Yang (C.N.), Mills (R.L.) Conservation of isotopic spin
and isotopic gauge invariance. 
Physical Review 96,1,191-195.

\end{document}